\documentclass[twocolumn,tighten]{aastex63}

\usepackage{amsmath}

\shorttitle{A Universal Acceleration Scale in Elliptical Galaxies}
\shortauthors{Chae et al.}

\begin{document}

\title{On the Presence of a Universal Acceleration Scale in Elliptical Galaxies}

\correspondingauthor{Kyu-Hyun Chae}
\email{chae@sejong.ac.kr, kyuhyunchae@gmail.com}

\author{Kyu-Hyun Chae}
\affil{Department of Physics and Astronomy, Sejong University, 
 209 Neungdong-ro Gwangjin-gu, Seoul 05006, Republic of Korea}

\author{Mariangela Bernardi}
\affil{Department of Physics and Astronomy, University of Pennsylvania,
209 South 33rd Street, Philadelphia, PA 19104, USA}

\author{Helena Dom\'{i}nguez S\'{a}nchez}
\affil{Department of Physics and Astronomy, University of Pennsylvania,
209 South 33rd Street, Philadelphia, PA 19104, USA}
\affil{Institute of Space Sciences (ICE, CSIC), Campus UAB, Carrer de Magrans, E-08193 Barcelona, Spain}

\author{Ravi K. Sheth}
\affil{Department of Physics and Astronomy, University of Pennsylvania,
209 South 33rd Street, Philadelphia, PA 19104, USA}

\begin{abstract}
  Dark matter phenomena in rotationally supported galaxies exhibit a characteristic acceleration scale of $g_\dagger \approx 1.2\times 10^{-10}$~m~s$^{-2}$. Whether this acceleration is a manifestation of a universal scale, or merely an emergent property with an intrinsic scatter, has been debated in the literature. Here we investigate whether a universal acceleration scale exists in dispersion-supported galaxies using two uniform sets of integral field spectroscopy (IFS) data from SDSS-IV MaNGA and ATLAS$^{\rm 3D}$. We apply the spherical Jeans equation to 15 MaNGA and 4 ATLAS$^{\rm 3D}$ slow-rotator E0 (i.e., nearly spherical) galaxies. Velocity dispersion profiles for these galaxies are well determined with observational errors under control. Bayesian inference indicates that all 19 galaxies are consistent with a universal acceleration of $g_\dagger=1.5_{-0.6}^{+0.9}\times 10^{-10}$~m~s$^{-2}$. Moreover, all 387 data points from the radial bins of the velocity dispersion profiles are consistent with a universal relation between the radial acceleration traced by dynamics and that predicted by the observed distribution of baryons. This universality remains if we include 12 additional non-E0 slow-rotator elliptical galaxies from ATLAS$^{\rm 3D}$. Finally, the universal acceleration from MaNGA and ATLAS$^{\rm 3D}$ is consistent with that for rotationally supported galaxies, so our results support the view that dark matter phenomenology in galaxies involves a universal acceleration scale.
 \end{abstract}

\keywords{ Dark matter (353); Non-standard theories of gravity (1118); Elliptical galaxies (456);  Modified Newtonian dynamics (1069) }


% \footnote{\url{http://www.latex-project.org/}}

\section{Introduction} \label{sec:intro}

Historically, empirical laws such as Kepler's kinematic laws have played crucial roles in physics. For the study of galaxies, these include the Faber-Jackson \citep{FJ76} and Fundamental Plane \citep{DD87,Dre87} relations for pressure-supported early-type galaxies, and the Tully-Fischer \citep{TF77}, baryonic Tully-Fisher \citep{McG00}, central density \citep{Lel16b}, and radial acceleration relations (RAR) \citep{MLS16}, for rotationally supported late-type galaxies.

The RAR -- a relation between the observed radial (centripetal) acceleration $g_{\rm obs}$ and the expected Newtonian acceleration $g_{\rm bar}$ due to the observed distribution of baryonic matter, for rotationally supported galaxies -- is of particular interest as it exhibits a characteristic acceleration scale (denoted $g_\dagger$) for dark matter phenomenology. If this scale is truly universal among all galaxies, it would naturally correspond to the critical acceleration (denoted $a_0$) of the modified Newtonian dynamics (MOND) paradigm \citep{Mil83}. On the other hand, if there is an intrinsic galaxy-to-galaxy scatter in $g_\dagger$, MOND would be called into question and $g_\dagger$ must be an emergent property of the dark matter (DM) phenomenology and/or the physics of galaxy formation.

When \cite{MLS16} reported the RAR from 153 galaxies selected from the SPARC database \citep{Lel16a}, they noted that data points from individual galaxies scattered around a universal RAR of $g_\dagger \approx 1.2\times 10^{-10}$~m~s$^{-2}$ with a typical scatter of $\sim 0.1$~dex consistent with typical observational errors. \cite{Li18} carried out Bayesian modeling of individual SPARC galaxies and noted that allowing a broad range of $g_\dagger$ did not improve overall fit qualities compared with the case of fixing (or imposing a narrow range of) $g_\dagger$. Therefore, up to realistic uncertainties, rotation curves of all rotationally supported galaxies appeared to be consistent with a universal $g_\dagger$.

This view was challenged by several authors \citep{Rod18a,Rod18b,CZ19,Mar20} who carried out Bayesian modeling using the formal uncertainties of individual rotation curves (and using uninformative priors in some cases). However, \cite{Kro18} and \cite{McG18} have highlighted the uncertain nature of some formal errors and the issue of appropriate priors, and \cite{CAB20} further pointed out the issue of the potential model misspecification in Bayesian applications to galaxy rotation curves arising from problems in data and/or the model.

\cite{SC19} considered an order-of-magnitude larger collection of rotation curves including the SPARC galaxies as a subsample (the ``PROBES'' sample) and argued for an intrinsic scatter in the RAR. However, individual rotation curves of the PROBES sample are typically not as accurate and extended as those of SPARC galaxies because they largely come from one-dimensional long-slit H$\alpha$ spectroscopy, which does not allow for accurate estimates of kinematic inclinations or identification of warped disks and noncircular motions. In addition, \cite{SC19} considered stellar mass rather than total baryonic mass distributions. However, unlike the SPARC mass models based on homogeneous Spitzer photometry at 3.6$\mu$m, in which the stellar mass-to-light ratio ($M_\star/L$) is approximately constant (e.g., \citealt{Schom19}), the mass models of \cite{SC19} are based on heterogeneous photometry in different bands, in which variations in $M_\star/L$ are a major concern (e.g.\ \citealt{MS14}). Therefore, it is more challenging to test the intrinsic scatter of $g_\dagger$ with PROBES galaxies. 

To broaden the discussion of the universality or not of $g_\dagger$, here we consider fitting the RAR to individual {\em dispersion}-supported elliptical galaxies, which we select from uniform integral field spectroscopic (IFS) surveys: SDSS-IV MaNGA and ATLAS$^{\rm 3D}$.  This is attractive for a number of reasons. First, both kinematic and photometric data come from uniform observations while SPARC and PROBES are collections of heterogeneously derived rotation curves. This means that measurement uncertainties of stellar velocity dispersions and light distributions of our galaxies are better understood. Second, we select only nearly spherical, slow-rotator (SR) galaxies and model them with the spherical Jeans equation. Although this means we are left with relatively few objects, modeling complications and errors can be minimized because of the relative simplicity of spherical models. Third, the velocity dispersion profiles of our galaxies cover acceleration ranges that are slightly larger than $g_\dagger$.  Although this weakens the sensitivity to $g_\dagger$, it also makes the external field effect less of a concern (see Figure 1 of  \citealt{Chae20}). Finally, our constraints on $g_\dagger$ are independent of those from rotationally supported galaxies. Thus, our constraints on the universality of $g_\dagger$ in elliptical galaxies can be directly compared with those from rotationally supported galaxies.  

As our goal is to test whether a common acceleration scale is present in elliptical galaxies, we will carry out kinematic modeling in the framework of MOND. In \S \ref{sec:dat} we describe our methodology and data. We present our results in \S \ref{sec:res}, and discuss and conclude in \S \ref{sec:dis}.

\section{Data and Methodology} \label{sec:dat}

\subsection{Galaxy Samples}

As we will use the spherical model, we select galaxies that are morphologically round E0s (defined here to be $b/a > 0.9$ where $a$ and $b$ are the semimajor and semiminor axes of the light distribution) and kinematically SRs. Because reliable kinematic modeling requires accurate velocity dispersion distributions of good spatial resolution, we consider the MaNGA \citep{Bun15} and ATLAS$^{\rm 3D}$ \citep{Cap11} databases. The ATLAS$^{\rm 3D}$ database provides velocity dispersion distributions with good spatial resolution but the sample size is small (260); the MaNGA database (not complete at the time of this writing) will eventually provide $\sim 10,000$ galaxies but with poorer spatial resolution than ATLAS$^{\rm 3D}$ \citep{Bun15} because the objects are more distant.

Of the subsample of 24 disk-less (i.e.\ pure-bulge) ATLAS$^{\rm 3D}$ galaxies selected in \cite{CBS18,CBS19}, 4 are round SRs:  NGC 4486, 4636, 5846, and 6703.  Their velocity dispersion maps and light distributions are shown in \cite{CBS18} and references therein.

We select galaxies from the MaNGA DR15 catalog \citep{DR15} as follows. Based on the photometric and morphological properties presented by \cite{Fisch19}, we require (1) TType $< 0$ (select early-types), (2) $P_{\rm S0} <0.3$ (small probability of being S0 rather than elliptical), (3) B/T $> 0.7$ (bulge-dominates light), (4) $\varepsilon (\equiv 1-b/a)<0.1$ (image is round), (5) $n_{\rm Ser}>3$ (light is centrally concentrated), (6) $\lambda_{\rm e} < 0.08+\varepsilon /4$ \cite[ensure SRs:  see][]{Fisch19}.  Of 4672 MaNGA DR15 galaxies only 30 satisfy these stringent criteria. 

\begin{figure}
  \centering
  \includegraphics[width=1.\linewidth]{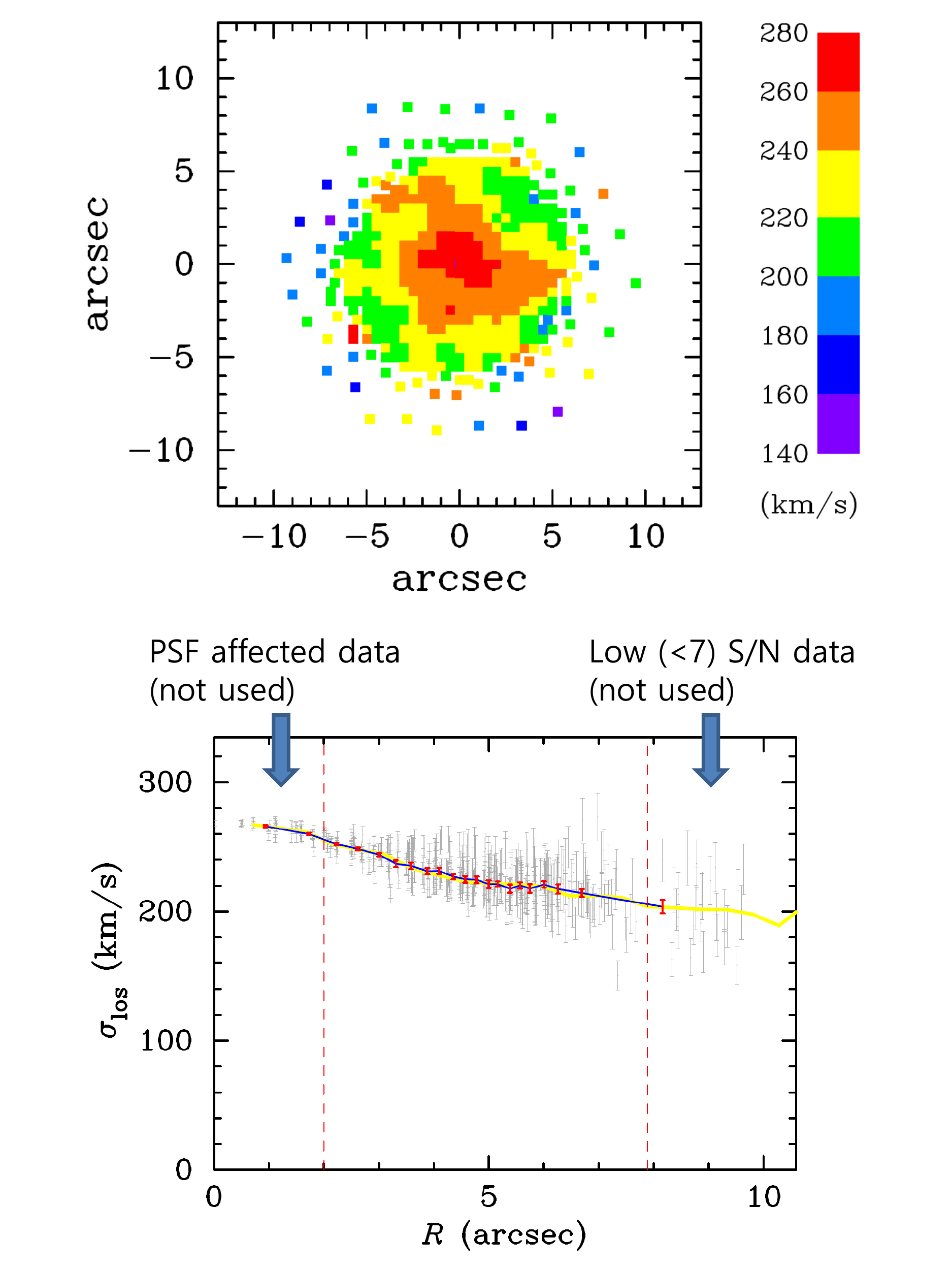}
    \vspace{-0.5truecm}  
    \caption{\small 
    Velocity dispersion map and the constructed radial profile of MaNGA IFU plate 9047-6102.
 (Top) Line-of-sight velocity dispersions measured at spaxels are represented by different colors. (Bottom) Radial bins are defined by concentric rings, and each bin except for the outermost one contains the same number of spaxels ($N=25$ in this case). The velocity dispersion in each ring is defined by the mean of the individual values and its uncertainty is defined by the standard deviation of the mean given by rms/$\sqrt{N-1}$. The inner $R<2''$ region is affected by the finite point spread function (PSF) size (a diameter of $2''$), and so not used. Also, low-S/N data in the outer region are not used as indicated.}
   \label{exam1}
\end{figure}

The velocity profiles of these galaxies are computed from the kinematic measurements derived by the MaNGA DAP contained in the MAPS-VOR10-GAU-MILESHC. An example is shown in Figure~\ref{exam1}. The velocity profile is constructed in two ways. The first one (the yellow line) is median value of the velocity dispersion of the unmasked spaxels with signal-to-noise (S/N)$>5$ in circularized radial bins of equal width. A $3\sigma$ clipping is also applied to prevent outliers from affecting the profile shape. The second one (the blue line and red points) is constructed by defining radial bins with an equal number of velocity dispersions ($N\ge 11$). Both agree and we use the red points with error bars for our modeling.

A MaNGA velocity dispersion map is discarded if it has too few independent values from which to construct a velocity dispersion radial profile with reasonable estimates of statistical uncertainties, or it is obviously asymmetric and therefore inconsistent with the spherical symmetry assumption, or the constructed radial profile does not cover at least twice the radius affected by the finite fiber size ($2''$).  Half of the selected MaNGA galaxies are discarded in this way, and thus we are left with 15 good-quality MaNGA spherical, SR galaxies. This cut based on IFU data quality will, of course, not be biased against any particular value of $g_\dagger$.

The selected 15 MaNGA SR E0s are listed in Table~1. These are all \emph{luminous giant ellipticals} brighter than a characteristic luminosity $L_\ast$ and larger than $\approx 5$~kpc.

\begin{table}[t]
\caption{List of MaNGA Selected Slow-rotator E0 Galaxies}
\begin{center}
  \begin{tabular}{ccccc} \hline
 IFU Plate  &  $M_r$ (mag)$^1$  &  $a$ (kpc)$^2$ & $n_{\rm Ser}$ & $b/a$ \\
 \hline
8243-6101 & $ -23.12 $ & $ 21.92 $ & $ 6.411 $ & $ 0.932 $  \\  
8249-6103 & $ -22.50 $ & $ 9.43 $ & $ 5.288 $ & $ 0.946 $  \\  
8323-9101 & $ -22.13 $ & $ 14.69 $ & $ 7.369 $ & $ 0.927 $  \\  
8459-6104 & $ -22.47 $ & $ 10.99 $ & $ 7.245 $ & $ 0.901 $  \\  
8482-1901 & $ -23.05 $ & $ 25.36 $ & $ 8.000 $ & $ 0.908 $  \\  
8555-6101 & $ -23.26 $ & $ 11.98 $ & $ 4.873 $ & $ 0.953 $  \\  
8718-3704 & $ -23.29 $ & $ 17.98 $ & $ 8.000 $ & $ 0.902 $  \\  
8726-9102 & $ -22.96 $ & $ 19.10 $ & $ 8.000 $ & $ 0.928 $  \\  
9024-3701 & $ -21.17 $ & $ 4.74 $ & $ 6.003 $ & $ 0.978 $  \\  
9047-6102 & $ -22.53 $ & $ 13.91 $ & $ 6.965 $ & $ 0.914 $  \\  
9088-3701 & $ -22.26 $ & $ 8.26 $ & $ 8.000 $ & $ 0.923 $  \\  
9485-6102 & $ -23.32 $ & $ 13.29 $ & $ 5.498 $ & $ 0.928 $  \\  
9501-6104 & $ -23.08 $ & $ 16.20 $ & $ 5.080 $ & $ 0.947 $  \\  
9868-3704 & $ -22.25 $ & $ 6.79 $ & $ 5.997 $ & $ 0.906 $  \\  
9888-6102 & $ -23.13 $ & $ 10.18 $ & $ 4.000 $ & $ 0.930 $  \\  
 \hline
\end{tabular}
\end{center}
Note: $^1$ SDSS $r$-band absolute magnitude is from a Sersic profile truncated at $8R_{\rm e}$.
    $^2$ Major axis of the ellipse enclosing one-half of the total light of the Sersic profile assuming $H_0=70$ km s$^{-1}$ Mpc$^{-1}$, $\Omega_{\rm M0}=0.3$, and $\Omega_{\rm \Lambda 0}=0.7$. The circularized effective radius used for modeling is $R_e\equiv \sqrt{ab}$. 
\end{table}

\subsection{Methodology}
The observed two-dimensional map of line-of-sight velocity dispersions that defines a radial profile $\sigma_{\rm los}(R)$ provides all the dynamical constraints for our models.  Figure~\ref{exam1} shows an example of good-quality MaNGA velocity dispersions and the range of scales that is actually used for kinematic modeling.

In our modeling we allow for velocity dispersion anisotropies and stellar mass-to-light ratio ($M_\star/L$) radial gradients. We follow the modeling procedures used recently \citep{CBS18,Chae19}. All our galaxies have negligible amounts of cold gas in the IFU-probed regions, so throughout baryons mean stars. Here we describe briefly the essential elements.

We work in the MOND framework in which the empirical\footnote{In rotationally supported galaxies the observed circular velocity is directly related to the centripetal acceleration via $g=V^2/R$. In dispersion-supported galaxies, the observed line-of-sight velocity dispersion is indirectly related to the radial acceleration via the spherical Jeans equation.} (i.e., dynamical) radial acceleration $g_{\rm dyn}(r)$ experienced by stars in a spherical system is related to the Newtonian acceleration due to baryons $g_{\rm bar}(r)$ as
\begin{equation}
g_{\rm dyn}(r) = \nu \left(\frac{g_{\rm bar}(r)}{g_\dagger}\right) g_{\rm bar}(r),
 \label{eq:RARmd}
\end{equation}
where $\nu(x)$ is known as the MOND interpolating function (IF) defining the RAR in the MOND framework, and $g_\dagger$ is the acceleration parameter of interest here. We consider 
\begin{equation}
 \nu(x) = \frac{1}{2}+\sqrt{\frac{1}{4}+\frac{1}{x}},
 \label{eq:IFnu}
\end{equation}
which is known as the Simple IF \citep{FB05}. This particular form seems to be preferred by elliptical galaxies \citep{Chae19}; it is similar to the form used by \cite{MLS16} to describe the SPARC galaxies, differing only in a subtle way at relatively high accelerations. Because we intend to test the universality of $g_\dagger$, we do not allow the IF functional form to vary. Our goal of testing the universality of $g_\dagger$ does not depend on this choice of the Simple IF. An alternative choice of the SPARC IF can shift the inferred $g_\dagger$ by an amount smaller than the uncertainty of the global value that we will obtain.

 We assume that the projected mass density of stars follows the observed surface brightness distribution with a possible radial variation of $\Upsilon_\star \equiv M_\star/L$ in the inner region $\la 0.8 R_{\rm e}$:
\begin{equation}
  \Upsilon_\star (R) =\Upsilon_{\star 0} \times \max\left\{1+ K\left[A - B (R/R_{\rm e})\right],1\right\},
 \label{eq:MLgrad}
\end{equation}
with $A=2.33$ and $B=3$ and $K$ is a parameter representing the strength of a gradient \citep{CBS18}. As estimated by \cite{Ber18} $K=1$ corresponds to the strong gradient reported in the literature \citep{vD17}, but here we take a mild gradient of $K=0.21$, i.e., $M_\star/L(R=0)=1.5\times M_\star/L(R\ge 0.8 R_{\rm e})$, with a scatter of $0.1$ based on \cite{DS19} and \cite{Ber19}. However, even if a broad range of $0<K<1$ is allowed with a flat prior as in \cite{Chae19}, it has only a minor impact on our study.

We can link the observed $\sigma_{\rm los}(R)$ profile with the empirical acceleration $g_{\rm dyn}(r)$ through the spherical Jeans equation \citep{BT08} and a velocity dispersion anisotropy profile $\beta(r) \equiv 1 - \sigma_{\rm t}^2(r)/\sigma_{\rm r}^2(r)$, where $\sigma_{\rm r}(r)$ and $\sigma_{\rm t}(r)$ are, respectively, the one-dimensional radial and tangential velocity dispersions. We assume a smoothly varying anisotropy profile of 
\begin{equation}
  \beta(r)=\beta_0 + (\beta_\infty - \beta_0) \frac{(r/r_a)^2}{1+(r/r_a)^2},
 \label{eq:gOM}
\end{equation}
where $\beta_0$ ($\beta_\infty$) is the anisotropy at $r=0$ ($\infty$) and $r_a$ is the radius where the anisotropy is the middle between the two.

Our model has the following free parameters:
$$\vec{\Theta}=\{M_{\star 0}(\equiv \Upsilon_{\star 0} L), K, \beta_0, \beta_\infty, r_a, g_\dagger \},$$
where $L$ is the luminosity. We impose the following priors: a Gaussian prior with $(\mu, \sigma)=(0.21,0.1)$ for $K$, and flat priors $-2 < \beta_0 < 0.7$, $-2 < \beta_\infty < 0.7$,  $0.1 R_{\rm e} < r_a < 1 R_{\rm e}$, and $-15 < \log_{10} g_\dagger < -5$ where $R_{\rm e}$ is the effective radius and $g_\dagger$ is in units of m~s$^{-2}$. In general Bayesian applications, it would be preferable to impose an empirical common prior distribution on $g_\dagger$ estimated from elliptical galaxies themselves. In our study, we carefully selected a clean sample of galaxies with well-determined $\sigma_{\rm los}$ profiles. This means that $g_\dagger$ may be robustly estimated in each galaxy even when the prior on its value is uninformative. The prior ranges on the other parameters are discussed in \cite{CBS18} and \cite{CBS19}.

We define a $\chi^2$ function by
\begin{equation}
  \chi^2 = \sum_{i=1}^{N_{\rm bin}} \left( \frac{\sigma_{\rm los}^{\rm mod}(R_i) -\sigma_{\rm los}^{\rm obs}(R_i)}{\delta_i} \right)^2  ,
 \label{eq:chi2}
\end{equation}
where $\sigma_{\rm los}^{\rm obs}(R_i)$ and $\delta_i$ are the mean and its error in each radial bin from the velocity dispersion map (see Figure~\ref{exam1}), and $\sigma_{\rm los}^{\rm mod}(R_i)$ is the MOND model prediction \citep{CBS18,Chae19}. The likelihood is $\propto e^{-\chi^2 /2}$ and the posterior probability density function (PDF) of $g_\dagger$ is derived from Markov Chain Monte Carlo (MCMC) simulations with the public code {\tt emcee} \citep{emcee}. As was discussed in the Appendix of \cite{Chae19}, because of the complexity of the parameter space a narrow distribution of initial walkers around the maximum-likelihood estimate of the parameters can produce unrealistically narrow PDFs. However, if initial walkers are widely distributed within the prior ranges, then the PDFs returned by MCMC widen and become more similar to the results from the simple Monte Carlo simulations discussed by \cite{Chae19}. Therefore, here we consider only MCMC simulations with widely sampled initial walkers, which are qualitatively similar to the simple Monte Carlo simulations.

\section{Results} \label{sec:res}

\begin{figure}
  \centering
  \includegraphics[width=1.0\linewidth]{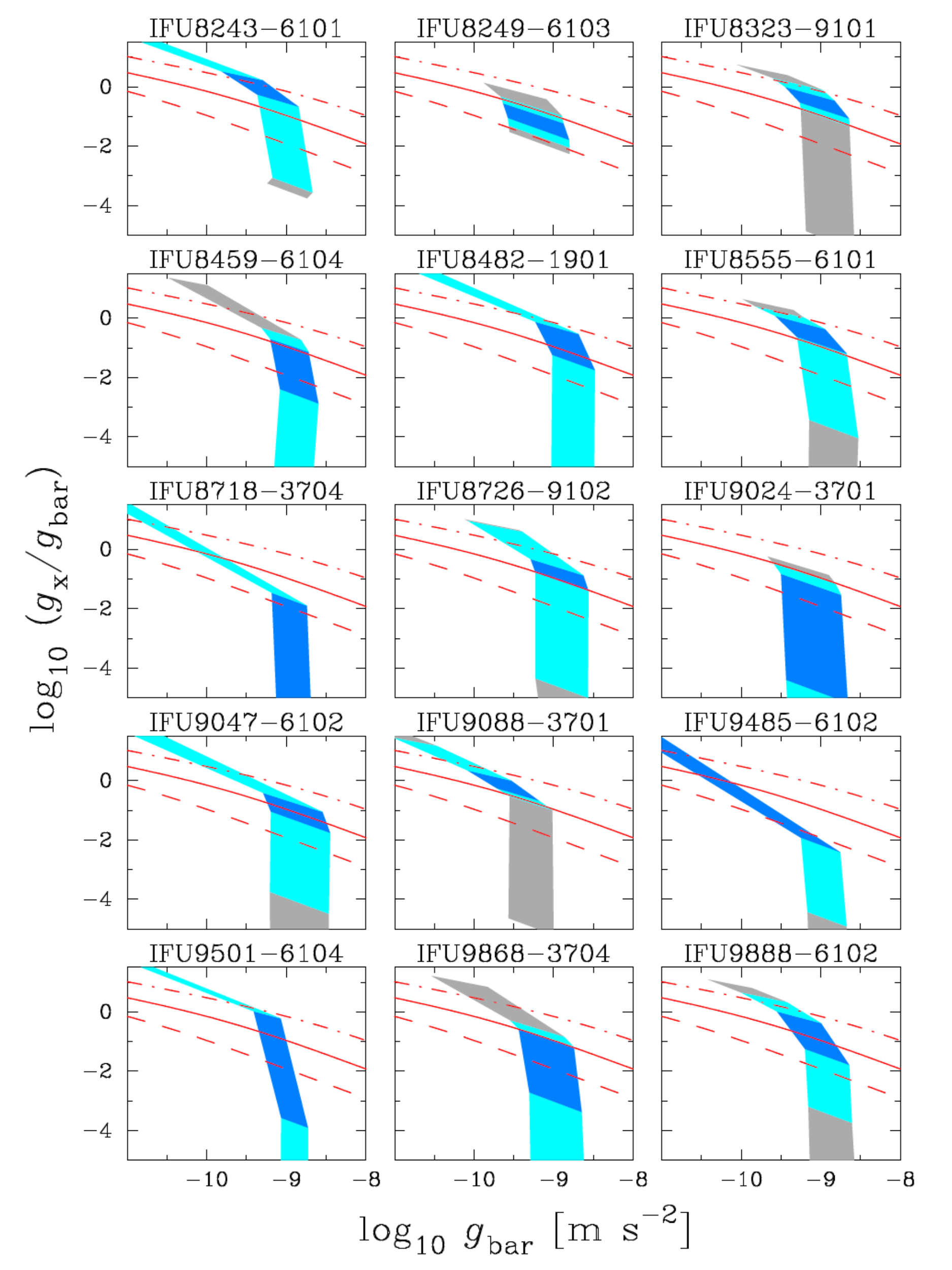}
    \vspace{-0.5truecm}  
    \caption{\small 
    MCMC results for MaNGA E0 SR galaxies.
    For galaxies with the IFU plate numbers given, the posterior probability distributions from MCMC outputs are shown in a modified RAR space by colored bands: blue $1\sigma$,  cyan $2\sigma$, and gray $3\sigma$. Here $g_{\rm x}\equiv g_{\rm dyn}-g_{\rm bar}$. The width of each band corresponds to a posterior range of $g_\dagger$. The red solid, dashed, and dash-dotted lines correspond to $g_\dagger = 1.2\times 10^{-10}$, $1.2\times 10^{-11}$, and $1.2\times 10^{-9}$ m~s$^{-2}$.}
   \label{gxgbar}
\end{figure}

Figure~\ref{gxgbar} shows MCMC results for 15 MaNGA E0 galaxies. It shows posterior probability distributions in the $g_{\rm x}/g_{\rm bar}$ (where $g_{\rm x}\equiv g_{\rm dyn}-g_{\rm bar}$) versus $g_{\rm bar}$ plane. Note that for the supercritical (i.e., $g_{\rm bar}\ga 10^{-10}$~m~s$^{-2}$) acceleration regime, subtly different cases can be better distinguished in this modified RAR space \citep{Chae19}. Similar results for ATLAS$^{\rm 3D}$ galaxies can be found in \cite{Chae19} and slightly revised results from this work are not shown. Figure~\ref{gxgbar} indicates that the case with $g_\dagger = 1.2\times 10^{-10}$~m~s$^{-2}$ is included within the $2\sigma$, i.e.\ 95 percent confidence limits (CLs) of every individual galaxy displayed.

\begin{figure}
  \centering
  \includegraphics[width=0.98\linewidth]{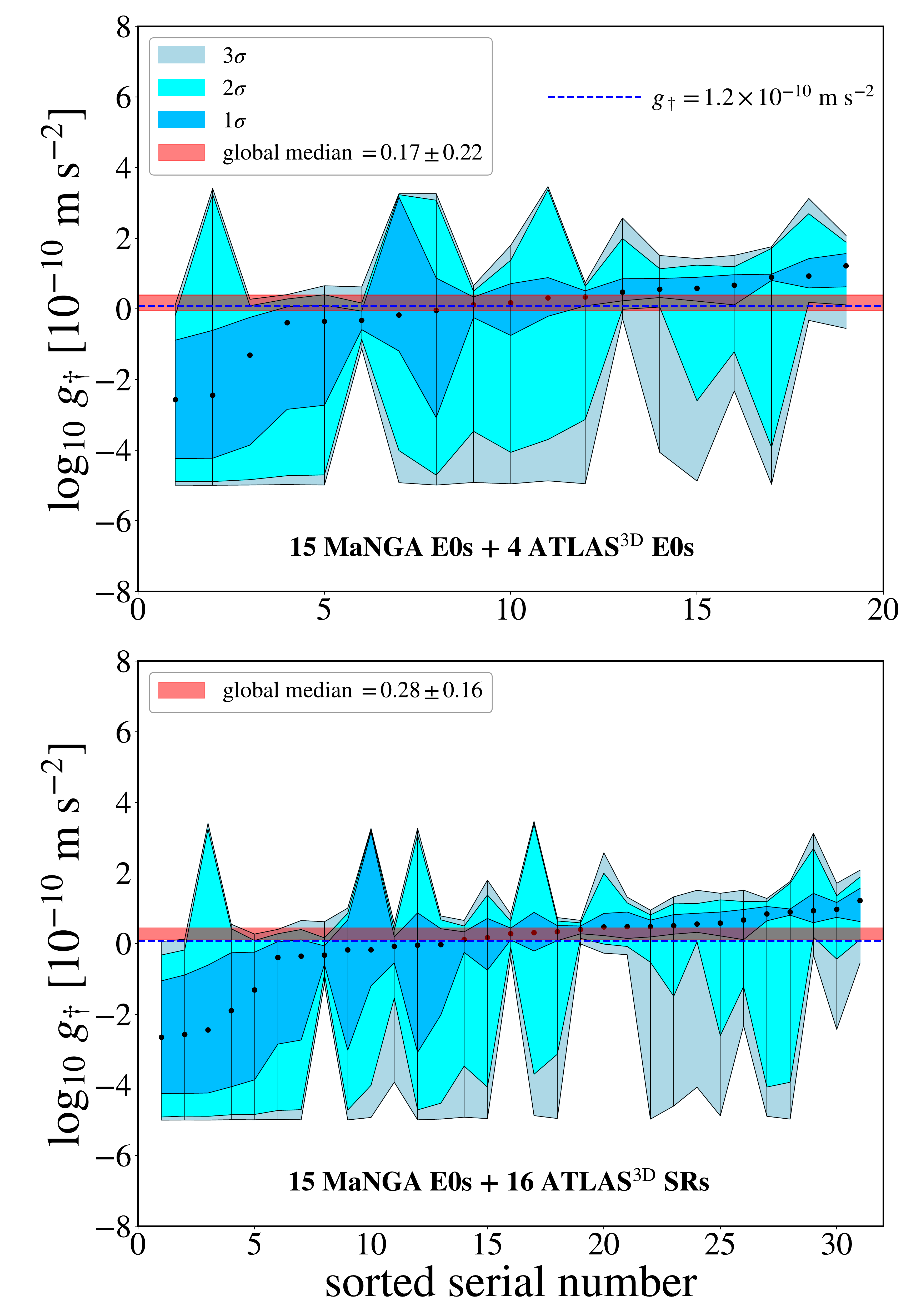}
    \vspace{-0.25truecm}  
    \caption{\small 
    Individually fitted $g_\dagger$ for elliptical galaxies.
 (Top) Individual PDFs of $g_\dagger$ for E0 SR galaxies selected from SDSS-IV MaNGA and ATLAS$^{\rm 3D}$. Individual PDFs are not Gaussian as can be seen by the confidence limits. The global median is defined by the median of the medians from the PDFs and its uncertainty is estimated by bootstrap resampling. All individual PDFs overlap so that all are consistent with the global median within $2\sigma$. (Bottom) This includes 12 ATLAS$^{\rm 3D}$ non-E0 SRs. The overall statistical properties are similar to the top panel.} 
   \label{g0all}
\end{figure}

The top panel of Figure~\ref{g0all} shows the individual PDFs of $g_\dagger$ for the 19 MaNGA and ATLAS$^{\rm 3D}$ E0 SRs sorted in order of ascending $g_\dagger$. As indicated in Figure~\ref{gxgbar}, individual PDFs are quite broad, with $2\sigma$ CLs covering $\sim 3$~dex. This is, in part, because uninformative priors are used for $g_\dagger$, but reflects the fact that our velocity dispersion profiles probe an acceleration range that is only weakly sensitive to $g_\dagger$.

It is evident from Figure~\ref{g0all} that all PDFs overlap one another well. All PDFs include the global median within $2\sigma$. Also, 12 out of 19 galaxies include the global median within $1\sigma$, i.e.\ 68 percent CL. Thus, our results are statistically consistent with a universal value of $g_\dagger$ among all E0 galaxies. The global median for our E0s $g_\dagger = 1.5_{-0.6}^{+0.9}\times 10^{-10}$ m~s$^{-2}$ is in excellent agreement with the canonical value for rotationally supported galaxies $g_\dagger=1.2\times 10^{-10}$ m~s$^{-2}$.

As a self-consistency check of our MCMC modeling we have considered other objects for which the spherical model is less well-motivated or inadequate. We first consider an extended sample which includes all 16 SRs from the ATLAS$^{\rm 3D}$ photometric pure-bulge sample \citep{CBS18} (i.e. we relax the criterion on the circularity of the light distribution). The bottom panel of Figure~\ref{g0all} shows the individual PDFs for all 31 SRs. Evidently, even non-E0 (i.e.\ not perfectly round) SRs are consistent with a universal value of $g_\dagger$. This may imply that the spherical model with velocity dispersion anisotropy can adequately describe nearly spherical SRs. If we include 8 fast-rotator (FR) ATLAS$^{\rm 3D}$ galaxies from the sample defined by \cite{CBS18}, then we find two individual results that deviate by more than  $3\sigma$ from the universal $g_\dagger$ value. Of course, this is not surprising because the spherical model is bound to fail for FRs; rather, it demonstrates self-consistency of our MCMC modeling. 

\begin{figure}
  \centering
  \includegraphics[width=1.0\linewidth]{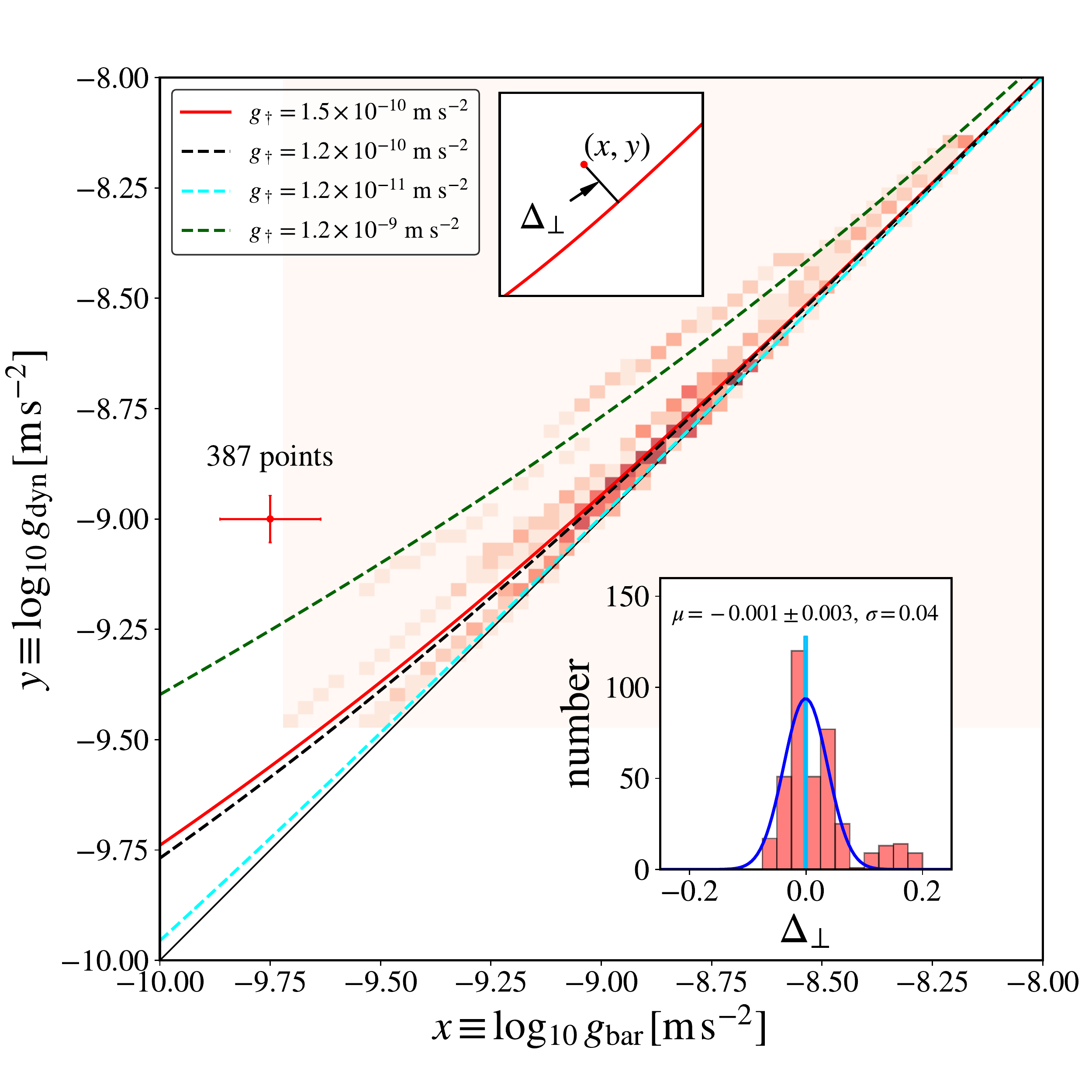}
    \vspace{-0.5truecm}  
    \caption{\small 
    RAR for MaNGA and ATLAS$^{\rm 3D}$ E0 SR galaxies.
    Heatmap shows $g_{\rm dyn}$ and $g_{\rm bar}$ for 387 data points from the 19 objects in the top panel of Figure~\ref{g0all}.  Error bar on the left of the plot shows typical uncertainties from the MCMC results.  Curves show Equation~(\ref{eq:RARmd}) for various $g_\dagger$ (as labeled).  Histogram in bottom right inset panel shows the distribution of offsets from the red curve; blue curve shows a Gaussian with rms 0.04.}
   \label{rar}
\end{figure}

Figure~\ref{rar} shows 387 data points from the MCMC analysis of the velocity dispersion profiles of the 19 E0s in the top panel of Fig.~\ref{g0all} in a format that is similar to the RAR for rotationally supported SPARC galaxies \citep{MLS16,Lel17}. Two tracks from MaNGA IFU~8243-6101 and NGC~4636 appear offset toward larger $g_\dagger$, but these points are actually consistent with the RAR curve within their $2\sigma$ uncertainties. These galaxies are not otherwise unusual. Overall, the inset shows that all 387 points are statistically consistent with a universal RAR, confirming the results shown in Figures~\ref{gxgbar} and \ref{g0all}.

\section{Discussion} \label{sec:dis}

In light of the recent debate on the presence of a universal acceleration scale involving dark matter phenomena in rotationally supported galaxies \citep{Kro18,McG18,Rod18a,Rod18b,CZ19,Mar20}, we have carried out kinematic analyses of a well-defined sample of E0 SR galaxies. Our Bayesian inference analysis yields individual posterior PDFs of $g_\dagger$ for 19 E0s that are consistent with a universal value of $g_\dagger=1.5_{-0.6}^{+0.9} \times 10^{-10}$~m~s$^{-2}$. This value agrees well with that reported by \cite{MLS16} for 153 rotationally supported galaxies from the SPARC database:  $g_\dagger=(1.2\pm 0.2)\times 10^{-10}$~m~s$^{-2}$ (including systematic error).

Two crucial aspects of our Bayesian methodology are (1) the selection and use of reliable kinematic data and (2) robust and wide searches of the parameter space within the prior ranges. When these requirements are met, the presence of a common parameter may be tested for statistically. When the estimated error models are uncertain and/or the kinematic model is inappropriate, Bayesian inference is not suitable for statistically testing a common parameter \cite[see also the discussion in][]{CAB20}. 

To test the universality of $g_\dagger$ in generic early-type galaxy samples, one {\em must} use accurate nonspherical models. The larger parameter space of such models, and the greater likelihood of parameter degeneracies, will make this challenging.  We believe that our results for (very nearly) spherical galaxies, which support the presence of a universal $g_\dagger$ for dark matter phenomena in galaxies, motivate future efforts to test the universality of $g_\dagger$ in generic galaxy samples, however challenging.

The presence of a universal acceleration scale fits naturally into the MOND paradigm: the empirical $g_\dagger$ becomes Milgrom's critical acceleration $a_0$.  We would have falsified MOND if we had not found a universal value, or if this value had been different from that for rotating galaxies. In this respect, we note that MOND analyses were carried out recently based on MaNGA \citep{Dur18} and ATLAS$^{\rm 3D}$ \citep{Tor14} galaxies, although our sample is more carefully defined and our analysis, which accounts for $M_\star/L$ radial gradients, is more general. Currently available $\Lambda$CDM simulations of galaxy formation and evolution predict intrinsic scatter of $\sim$0.06-0.08~dex in the value of $g_\dagger$ around the empirical mean value (see \cite{SC19} and references therein, in particular, \cite{Dut19}). As this is substantially smaller than our current error bars, larger samples are needed to test if this is consistent with the data.

%\vspace{0.3in}

\acknowledgements
We thank F. Lelli, S. McGaugh, and P. Li for discussion and useful comments on the manuscript. We thank the anonymous reviewer for useful comments that helped us improve the presentation.
 This research was supported by the National Research Foundation of Korea (NRF) grant funded by the Korea government (MSIT) (No.\ NRF-2019R1F1A1062477). M.B. acknowledges support from the NFS grant AST-1816330.
H.D.S. acknowledges support from the Centro Superior de Investigaciones Científicas PIE2018-50E099.


\begin{thebibliography}{}

\bibitem[Aguado et al.(2019)]{DR15} Aguado, D. S., Ahumada, R., Almeida, A., et al.\ 2019, \apjs, 240, 23

\bibitem[Bernardi et al.(2019)]{Ber19}  Bernardi, M., Dom\'{i}nguez S\'{a}nchez, H.,  Brownstein, J. R., Drory, N.,  Sheth,  R. K. 2019, \mnras, 489, 5633
  
\bibitem[Bernardi et al.(2018)]{Ber18} Bernardi, M., Sheth, R. K., Dominguez-Sanchez, H., et al.\ 2018, MNRAS, 477, 2560
  
\bibitem[Binney \& Tremaine(2008)]{BT08} Binney, J., Tremaine, S. 2008, Galactic Dynamics, (2nd ed.; Princeton, NJ: Princeton Univ.\ Press)

\bibitem[Bundy et al.(2015)]{Bun15} Bundy, K., et al.\ 2015, \apj, 798, 7

\bibitem[Cameron, Angus \& Burgess(2020)]{CAB20} Cameron, E., Angus, G. W., Burgess, J. M. 2020, NatAs, 4, 132  

\bibitem[Cappellari et al.(2011)]{Cap11} Cappellari, M., Emsellem, E., Krajnovi\'{c}, D., et al.\ 2011, \mnras, 413, 813
    
\bibitem[Chae, Bernardi \& Sheth(2018)]{CBS18} Chae, K.-H., Bernardi, M., Sheth,  R. K. 2018, \apj, 860, 81

\bibitem[Chae, Bernardi \& Sheth(2019)]{CBS19} Chae, K.-H., Bernardi, M., Sheth,  R. K. 2019, \apj, 874, 41
  
\bibitem[Chae et al.(2019)]{Chae19} Chae, K.-H., Bernardi, M., Sheth,  R. K., Gong, I.-T. 2019, \apj, 877, 18
  
\bibitem[Chae et al.(2020)]{Chae20} Chae, K.-H., Lelli, F., Desmond, H., McGaugh, S. S., Li, P., Schombert, J. M. 2020, ApJ, in press (arXiv:2009.11525)
  
\bibitem[Chang \& Zhou(2019)]{CZ19} Chang, Z., Zhou, Y. 2019, \mnras, 486, 1658
  
\bibitem[Djorgovski \& Davis(1987)]{DD87} Djorgovski, S., Davis, M. 1987, \apj, 313, 59

\bibitem[Dom\'{i}nguez S\'{a}nchez et al.(2019)]{DS19}  Dom\'{i}nguez S\'{a}nchez, H., Bernardi, M., Brownstein, J. R., Drory, N.,  Sheth,  R. K. 2019, \mnras, 489, 5612
  
\bibitem[Dressler et al.(1987)]{Dre87} Dressler, A., Lynden-Bell, D., Burstein, D., Davies, R. L., Faber, S. M., Terlevich, R., Wegner, G. 1987, \apj, 313, 42

\bibitem[Durazo et al.(2018)]{Dur18} Durazo, R., Hernandez, X., Cervantes Sodi, B., \& Sanchez, S. F. 2018, \apj,
863, 107

\bibitem[Dutton et al.(2019)]{Dut19} Dutton, A. A., Macci\`{o}, A. V., Obreja, A., Buck, T. 2019, \mnras, 485, 1886
  
\bibitem[Faber \& Jackson(1976)]{FJ76} Faber, S. M., Jackson, R. E. 1976, \apj, 204, 668
  
\bibitem[Famaey \& Binney(2005)]{FB05} Famaey, B., Binney, J. 2005, \mnras, 363, 603

\bibitem[Fischer et al.(2019)]{Fisch19} Fischer, J.-L., Dominguez-Sanchez, H., Bernardi, M. 2019, \mnras, 483, 2057

\bibitem[Foreman-Mackey et al.(2013)]{emcee} Foreman-Mackey, D., Hogg, D. W., Lang, D., Goodman, J. 2013, \pasp, 125, 306 
  
\bibitem[Kroupa et al.(2018)]{Kro18} Kroupa, P., Banik, I., Haghi, H., Zonoozi, A., et al. 2018, NatAs, 2, 925

\bibitem[Lelli, McGaugh \& Schombert(2016)]{Lel16a} {{Lelli}, F.} , {{McGaugh}, S.~S.}, {{Schombert}, J.~M.} 2016, {\aj}, {152}, {157}

\bibitem[Lelli et al.(2016)]{Lel16b} {{Lelli}, F.} , {{McGaugh}, S.~S.}, {{Schombert}, J.~M.}, {{Pawlowski}, M.~S.} 2016, {\apjl}, {827}, {L19}

\bibitem[Lelli et al.(2017)]{Lel17} Lelli, F., McGaugh, S. S., Schombert, J. M., Pawlowski, M. S. 2017, \apj, 836, 152

\bibitem[Li et al.(2018)]{Li18} Li, P., Lelli, F., McGaugh, S., Schombert, J. 2018, \aap, 615, 70  
  
\bibitem[Marra et al.(2020)]{Mar20} Marra, V., Rodrigues, D. C., de Almeida, A. O. F. 2020, \mnras, 494, 2875
  
\bibitem[McGaugh, Lelli \& Schombert(2016)]{MLS16} McGaugh, S. S., Lelli, F., Schombert, J. M. 2016, \prl, 117, 201101

\bibitem[McGaugh et al.(2018)]{McG18} McGaugh, S. S., Li, P., Lelli, F., Schombert, J. M. 2018, NatAs, 2, 924

\bibitem[McGaugh \& Schombert(2014)]{MS14} McGaugh, S. S., Schombert, J. M. 2014, \aj, 148, 77
  
\bibitem[McGaugh et al.(2000)]{McG00} {{McGaugh}, S.~S.}, {{Schombert}, J.~M.}, {{Bothun}, G.~D.}, {{de Blok}, W.~J.~G.} 2000, {\apjl}, {533}, {L99}

\bibitem[Milgrom(1983)]{Mil83} Milgrom, M. 1983, \apj, 270, 371

\bibitem[Rodrigues et al.(2018a)]{Rod18a} Rodrigues, D. C., Marra, V., del Popolo, A., Davari, Z. 2018a, NatAs, 2, 668

\bibitem[Rodrigues et al.(2018b)]{Rod18b} Rodrigues, D. C., Marra, V., del Popolo, A., Davari, Z. 2018b, NatAs, 2, 927

\bibitem[Schombert, McGaugh \& Lelli(2019)]{Schom19} {{Schombert}, J.~M.}, {{McGaugh}, S.~S.} \& {{Lelli}, F.} 2019, {\mnras}, {483}, {1496}

\bibitem[Stone \& Courteau(2019)]{SC19} {{Stone}, C.}, {{Courteau}, S.} 2019, {\apj}, {882}, {6} 

\bibitem[Tortora et al.(2014)]{Tor14} Tortora, C., Romanowsky, A. J., Cardone, V. F., Napolitano, N. R., \& Jetzer, Ph. 2014, MNRAS, 438, L46
  
\bibitem[Tully \& Fisher(1977)]{TF77} Tully, R. B., Fisher, J. R. 1977, \aap, 54, 661

\bibitem[van Dokkum et al.(2017)]{vD17} van Dokkum, P., Conroy, C., Villaume, A. , Brodie, J., Romanowsky, A. J. 2017, \apj, 841, 68 

  
\end{thebibliography}
\end{document}